\documentclass[12pt]{iopart}
\usepackage{iopams}
\usepackage[latin1]{inputenc}
\usepackage {setstack}
\usepackage{amsfonts}
\usepackage{amssymb}
\usepackage{graphicx}
\usepackage{subfigure}
\usepackage{float}
\usepackage{cite}

\begin{document}

\author{F V Day \footnote {Present Address: St. Catharine's College, Trumpington Street, Cambridge, CB2 1RL} and S D Barrett}
\address{ Quantum Information Theory Group, Level 12, Electrical Engineering Building, Imperial College London, Exhibition Road, London SW7 2AZ}
\eads{\mailto{francescavday@gmail.com}, \mailto{seandbarrett@gmail.com}}
\title{The Ising ferromagnet as a self-correcting physical memory: a Monte-Carlo study}


\begin{abstract}

The advent of quantum computing has heralded a renewed interest in
physical memories - physically realizable structures that offer
reliable data storage with error correction only at the point of
access. Here, we examine a model of a classical physical memory,
capable of storing a classical bit, based on the ferromagnetic Ising
model on 1- and 2-dimensional square lattices. We make use of
Monte-Carlo simulations as well as analytic solutions in the high
temperature limit. At high temperatures, both 1- and 2-D Ising models
behave like a non-interacting model, albeit with a reduced effective
number of spins and a reduced effective spin-flip rate. In agreement
with general arguments, we confirm numerically that (i) the
information storage time is independent of system size in 1D, (ii) in
two dimensions, at temperatures below the Onsager phase transition
temperature $T_{c}$, the information storage time increases exponentially
with system size. Interestingly, some benefit in increasing system
size is found even above $T_{c}$, although the increase in lifetime is not
exponential.

\end{abstract}

\maketitle

\section{Introduction}
The advent of quantum computing\cite{QC} has heralded a renewed interest in self-correcting physical memories - physically realizable structures that offer reliable data storage with error correction only at the point of access. Physical storage of quantum bits presents many challenges, and model memory structures such as the surface code Hamiltonian\cite{toric} have been devised to meet these. In classical computing, by contrast, Nature has provided us with a physical memory: the ferromagnet, which is the basis of magnetic data storage\cite{storage}. This article investigates quantitatively the use of the ferromagnet as a physical memory. We hope that the understanding gained of this simple case will also provide a starting point for a deeper understanding of quantum memories.\\

The Ising model, proposed by Lenz in 1920\cite{Lenz}, considers an anisotropic ferromagnet as a $D$-dimensional lattice of spins, each of which may take values $s_{i} = \{1,-1\}$. The energy of this Ising lattice is given by:
\begin{equation}
H = -J\sum_{<i,j>} s_{i}s_{j} - \sum_{i} h_{i}s_{i}
\end{equation} where the first sum is over nearest neighbour pairs. $J$ represents the strength of the magnetic interaction between neighbouring spins, while $h$ is the external magnetic field. For a ferromagnet, $J>0$. In this work, we take $h=0$, $J=1$. For $h=0$, $H$ is minimised when all spins are equal, so the Ising lattice has two energy minima.

Initial investigations asked whether or not an infinite Ising lattice displays a phase transition. This would correspond to the transition from para- to ferro- magnetism as the temperature of a magnetic material is decreased. Ising\cite{Ising} showed that there is no phase transition for $D=1$. Kramers and Wannier\cite{KW1,KW2} later showed that a phase transition does occur for $D=2$. Onsager\cite{Onsager} identified the critical temperature at which this transition occurs as:
\begin{equation}
\tanh\frac{2J}{kT_{c}} = \frac{1}{\sqrt{2}} \Rightarrow \frac{kT_{c}}{J} \backsimeq 2.269185
\end{equation}
The critical temperature is equivalent to the Curie temperature of a magnetic material: for $T<T_{c}$ spontaneous magnetization occurs. The magnitude of the spontaneous magnetization was calculated exactly by Yang\cite{Yang} in 1952. \\

The energy landscape of an Ising lattice is the basis of classical magnetic data storage. We model the storage process by attempting to store a single bit, $x$, using an Ising lattice of $N$ spins with side length $L$. Each energy minimum corresponds to a different value of the bit. For example, we might choose:
\begin{equation}
\eqalign {s_{i} = -1 \forall i \Longleftrightarrow x = 0 \cr
s_{i} = 1 \forall i \Longleftrightarrow x = 1}
\end{equation}
Thermal spin fluctuations may corrupt our lattice. Fortunately, these will carry an energy penalty, so the system will tend to correct such fluctuations. In this way, the Ising lattice acts as a self-correcting physical memory. After some time, we attempt to retrieve our stored bit by majority voting. This corresponds to a final error correction step at the point of readout. Such a step is always necessary when evaluating the performance of a self-correcting memory based on an error correcting code, as has been noted by Pastawski et al. \cite{Pastawski}. For example, if more than half\footnote {In the unlikely event that exactly half of the spins have the value -1 we have two options: (1) Choose the value of $x$ randomly or (2) Declare that the data has been corrupted. The latter is mathematically equivalently to inferring the wrong value of the bit: in either case, the memory has failed. We will use whichever of these is mathematically convenient.} of the spins have the value -1 we infer that $x = 0$. From an information-theoretic viewpoint, the Ising model is a physical realisation of a repetition code. Likewise, the surface code Hamiltonian represents a physical realisation of a \emph{quantum} error correcting code.\\

The energy penalty during the evolution of the system occurs on the perimeter of regions of different spin. For a 1D Ising lattice, a region of incorrect spin may therefore grow without further energy penalty. However, growth of incorrect spin regions in lattices with $D>1$ requires further energy input from the environment. This argument, proposed by Peierls \cite{Peierls}, leads one to expect that the 1D Ising lattice is a poor physical memory, whereas lattices with $D>1$ are good physical memories at sufficiently low temperature. One aim of this paper is to explore quantitatively these ideas via Monte-Carlo (MC) simulations.

An analogous effect is seen in quantum physical memories, in particular in the toric codes described in reference \cite{toric}. In a 2D toric code, the logical operators that take the system from one ground state to another are string-like. This means that, once created, quasi-particles representing the propagation of these errors can travel without energy penalty. Consequently, the 2D toric code is not a good physical memory. In the 4D toric code Hamiltonian, by contrast, logical operators are membrane-like, and therefore the propagation of errors is accompanied by an additional energy penalty, and a good physical memory is expected to result \cite{toric,stability}. Recently, a 3D stabilizer code Hamiltonian with no string-like logical operators has been proposed, which does energetically penalise additional errors and therefore may serve as a useful quantum information storage device\cite{Haah,Haah2,Haah3}.\\

We define the fidelity $F(t$) of a memory as the probability that the correct bit will be retrieved at time $t$, given that the bit was encoded perfectly at $t = 0$. For an Ising lattice, $F(t)$ is expected to depend on dimensionality, geometry and temperature. In this article, we will develop a quantitative model for $F(t)$ for Ising lattices with free boundaries and no external field. We will use MC simulations to verify the model and investigate its implications for 1D and 2D lattices at various temperatures.

We observe two distinct mechanisms of fidelity loss in an Ising lattice, depending on its temperature. For very high $T$, the lattice will decay into a disordered state, with no large regions of equal spin. For very low $T$, the lattice will suffer only small fluctuations from its initial state for a long period of time, until the unlikely event that a fluctuation grows to cover the whole of the lattice. The lattice will then remain in this new state, with the vast majority of spins incorrect, until another such fluctuation reverts the lattice to its original state. For intermediate $T$, at sufficiently large times the lattice is composed of a few regions of equal spin. These ideas are seen also in work on the kinetics of an Ising lattice in an opposing magnetic field (i.e. with spins initially aligned against the field). This area has benefited from extensive research for a number of years, see for example references \cite{D1,D2,D3,D4,D5,D6,D7,grains,D9,PhysMem,D11,D12,D13,D14,BCs}. In this environment, droplets of spin aligned with the field will form, but due to the energy penalty at the perimeter small droplets are likely to shrink. However, if a droplet reaches a critical size it will grow to fill the entire lattice. When $h\neq0$, the lattice will likely remain in this configuration (the minimum energy state) for an extremely long time. In our case, the lattice has a free choice of energy minimum, and so will continue to behave stochastically.

Ising lattice fidelity\footnote{This work studies the probability that the sign of the magnetization remains unchanged, as this is measurable with magnetic force microscopy experiments. This is equivalent to fidelity as defined here.} was studied using MC simulations by Kolesik et al\cite{grains} for a lattice with $D = 2$ at temperatures $T<T_{c}$ in an opposing field with periodic boundary conditions. The Ising lattice with free boundaries in an opposing field has also been investigated, for example by Cirillo and Lebowitz \cite{PhysMem}, but the fidelity is not considered in this work. Cirillo and Lebowitz demonstrate that the use of free rather than periodic boundaries is highly significant to the dynamics of the Ising lattice. With free boundary conditions, the critical droplet size, and hence the lifetime of the metastable state, is lower. It was also noted that the critical droplet develops in a corner of the system, where the energy of the perimeter is lower. Real ferromagnets, although they may display an exceedingly large value of $N$, have free boundaries.\\

In the next section we consider the dynamics in the limiting cases of high and low $T$. We then describe our procedure for simulating the time evolution of the fidelity in the case of arbitrary temperatures, and propose a model function for this time evolution, inspired by the high-temperature limiting behaviour. In section 3 we present and discuss the results of our Monte-Carlo simulations. In section 4 we conclude and point out some questions for future study.

\section{Dynamics}
\subsection{Limiting Cases}

In general, the evolution of the lattice under the influence of thermal noise will depend on the details of the system-bath interaction. In order to make progress, we do not assume a specific form of this interaction, but rather model the noise as a stochastic process that acts locally to flip spins and yields a long-time solution that is consistent with a Boltzmann distribution at temperature $T$. We first consider the limiting cases of very high and very low temperature. We begin our investigation by considering $N$ non-interacting spins. This is the limit as $T\rightarrow\infty$ of any Ising lattice with $N$ spins. At $t = 0$, we encode $x = 1$ i.e. $s_{i} = 1 \forall i$. At some later time, the spin $s_{i}$ has changed value $n_{i}$ times, where $n_{i}$ is modelled by a Poisson distribution with some characteristic flipping rate $\lambda$ per unit time. If n is even, $s_{i} = 1$; if n is odd, $s_{i} = -1$. Therefore:
\begin{equation}
P(s_{i} = -1)  = \sum_{n=0}^{\infty} \frac{\rme^{-\lambda t}(\lambda t)^{2n+1}}{(2n+1)!}= \frac{1}{2}(1-\rme^{-2\lambda t})
\end{equation}
The fidelity is the probability that fewer than half of the spins have flipped an odd number of times\footnote{Here we have chosen option (2) for the case of a 50/50 split.} which can be obtained using the binomial distribution:

\begin{equation}
F(t) = \frac{1}{2^{N}}\sum_{n=0}^{\lfloor N/2 \rfloor} \left( \begin{array}{c}
N\\
n
\end{array} \right) (1-\rme^{-2\lambda t})^{n}(1+\rme^{-2\lambda t})^{N-n}
\end{equation}

In the very low temperature case, the lattice is usually found in a state in which almost all spins are equal because the energy barrier between these minima is very high. As explained above, loss of fidelity is caused by the non-zero probability of the lattice transitioning between the minima. Therefore an approximate model for the low temperature case considers the whole lattice as a single spin with a very low $\lambda$. This phenomenon  is known as superparamagnetism\cite{neel} and presents a challenge for magnetic data storage\cite{spm1,spm2,spm3}. This simple interpretation gives an exponential form of the fidelity (equivalent to setting $N = 1$ in (5) ), as noted by Kolesik et al \cite{grains}.\\

\subsection{Monte-Carlo Simulations at Arbitrary Temperatures}

We used an MC process to simulate the evolution of the Ising lattice. At each MC time-step, we updated a single randomly chosen spin. We flipped the sign of the spin with a probability given by Glauber dynamics:

\begin {equation}
P(flip) = \frac{1}{1+\rme^{\frac{\Delta E}{kT}}}
\end {equation}
where $\Delta E = \rm{energy\: of\: lattice\: with\: spin\: flipped} - \rm{current\: energy\: of\: lattice}$. In order to make meaningful comparisons between systems of different sizes, physical time is modelled as $t = \frac{MC\:time-steps}{N}$. We approximated the fidelity by repeating each simulation $M = 10^{4}$ times and finding the fraction of these that encoded the correct value of $x$. In the non-interacting case, we have always $P(flip) = 0.5$, so each spin flips on average 0.5 times per physical time-step, and we expect $\lambda = 0.5$. At any given time, each of our $M$ runs has $F = 1$ or $F = 0$. Therefore $F(t)$ has a binomial distribution with $M$ trials and  probability of success $F(t)$, normalised by dividing by $M$. This allows us to find the statistical error in $F(t)$. Because $M$ is finite, we have also a discretization error. Errors in the fidelity calculated from the MC data were approximated by taking the largest of the discretization error and the statistical error:

\begin {equation}
\sigma_{F} = \rm{max} \left \lbrace \frac{1}{2M} \rm{,} \left [\frac{F(1-F)}{M} \right ]^{1/2} \right \rbrace
\end {equation}

MC simulations of the non-interacting case are an excellent fit to the analytical result, as shown in figure 1. As we discuss further in later sections of this paper, the basic form of the fidelity curve for an Ising lattice is a deformation of that for non-interacting spins. In particular, we note that, even at low temperature and in both dimensionalities simulated, the initial region of the fidelity curve with negative curvature (seen clearly in the non-interacting case) is still present. Therefore a simple exponential model cannot capture all the features of fidelity decay.

\begin{figure} [H]
\centering
\includegraphics[scale=0.25]{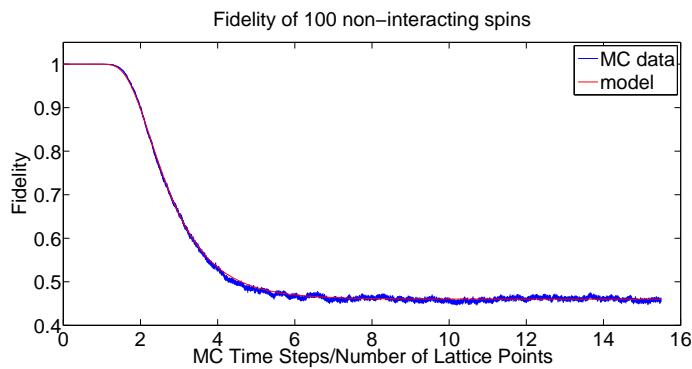}
\caption{MC data (blue with vertical error bars) and model (red) for the fidelity of 100 non-interacting spins. $\chi^{2} = 1554$; $\rm{Degrees\: of\: Freedom\:} = 1550$}
\end{figure}

\subsection {A Mathematical Model for $F(t)$}

Observations of the evolving Ising lattice verify that, in general, the lattice exhibits several spin regions (see figure 2) , which may move around the lattice and change sign. We therefore postulate that the Ising lattice can be approximately modelled using (5) by allowing $\lambda$ and the effective number of spins $N_{\rm{eff}}$ to vary. $N_{\rm{eff}}$ is interpreted as the average number of spin regions in the lattice.

\begin{figure} [H]
\centering
\includegraphics[scale=0.5]{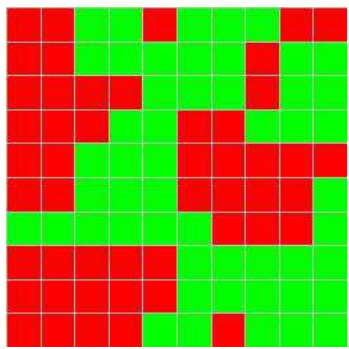}
\caption{A square Ising lattice with $N=100$ and $kT=2.5$ after 177000 MC steps. Initially all spins were +1 (red). Our model predicts $N_{\rm{eff}} = 5.37 \pm 0.09$ for this lattice.}
\end{figure}

To allow continuous variation of $N_{\rm{eff}}$, we used a Gaussian approximation\footnote{This models option (1) in the case of a 50/50 split.} for the binomial distribution. Our proposed model, then, is:

\begin {equation}
\eqalign{
F(t) = \frac{1}{2} \left (1+\rm{erf}\frac{\frac{N}{2}-\mu}{\sqrt{2}\sigma} \right ) \cr
\mu = \frac{1}{2}N_{\rm{eff}}(1-\rme^{-2\lambda t}) \cr
\sigma = \frac{1}{2}[N_{\rm{eff}}(1-\rme^{-2\lambda t})(1+\rme^{-2\lambda t})]^{1/2} }
\end {equation}
Figure 3 shows how the shape of $F(t)$ varies with $N_{\rm{eff}}$ and compares the Gaussian approximation to the exact binomial form. We can see that decreasing $N_{\rm{eff}}$ decreases the size of the flat region at low t with respect to the length of the decay, and overall quickens the decay of the fidelity. For a given lattice, however, a decrease in $N_{\rm{eff}}$ is accompanied by a decrease in $\lambda$, as the larger spins regions flip less frequently. A decrease in $\lambda$ preserves the shape of the curve, but stretches it along the time axis.

\begin{figure} [H]
\includegraphics[scale=0.3]{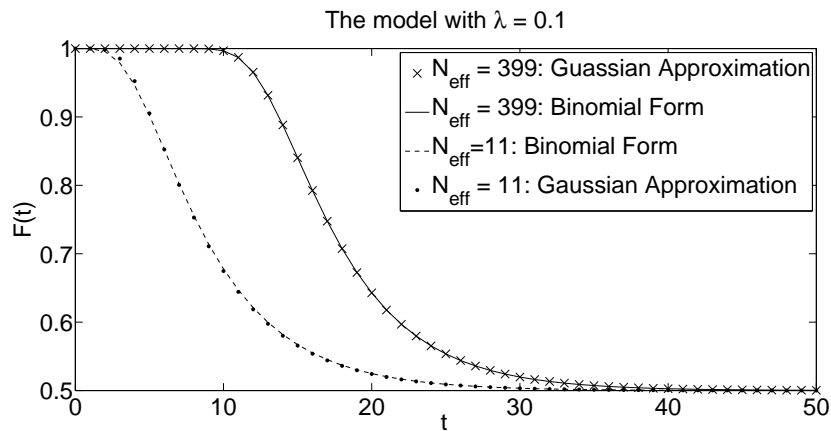}
\caption{Binomial (lines) and Gaussian (points) form of the model with $\lambda=0.1$; $N_{eff} = 399$ (solid line, crosses) and $N_{eff} = 11$ (dashed line, circles)}
\end{figure}

\section{Results and Discussion}

We simulated 1D and square 2D lattices with $N = \lbrace10^{2}, 11^{2},...,20^{2}\rbrace$. For the 1D lattices we used $kT = \lbrace 0.5, 1.0,...,8.0 \rbrace$; for the 2D lattices we used $kT = \lbrace 2.0, 2.1, ... 3.0 \rbrace $, which includes temperatures on either side of $T_{c}$. We also simulated a wider range of temperatures for $N = 100$. We used the non-linear least squares method to fit the MC data to the model, allowing $N_{\rm{eff}}$ and $\lambda$ to vary freely. Each data point was weighted by $\frac{1}{\sigma_{F}^{2}}$. The errors in $N_{\rm{eff}}$ and $\lambda$ are $90\%$ confidence intervals. For sufficiently high temperatures, the model gives a very good fit to the data (see figures 4 and 5). In fact, some of the $\chi^{2}$ values are significantly lower than expected, which is likely due to equation 7 giving an overestimate of the errors when $F$ is close to 1. However, as the temperature is decreased, the fit becomes poorer. This occurs at $T\simeq 0.3$ for $D = 1$ and at $T\simeq 2.1$ for $D = 2$, although the transition is not abrupt.  At temperatures close to this transition, $N_{\rm{eff}} <1 $ was often produced by the fitting, which is clearly unphysical. Additionally, the Gaussian approximation of the binomial distribution breaks down for the low values of $N_{\rm{eff}}$ needed at low $T$. In general, however, the fitting process gave $1 < N_{\rm{eff}} < N$ and $0 < \lambda <0.5$ as expected. The deviation of the actual lattice evolution from our model at low temperatures is not unexpected, as the model is extrapolated from the non-interacting case. However, we can see from figure 5(d) that a simple exponential model is also inadequate.
\\

We immediately notice that for low $T$ the decay of $F$ is significantly slower for $D=2$ than for $D=1$, when $N$ and $T$ are the same in each case (see figures 4 and 5). By considering the variation of the model parameters with $N$, a deeper insight into the differences between the 1D and 2D lattices may be achieved. (Variation of the model parameters with $T$ is discussed in the appendix.)

\begin{figure} [H]
\subfigure[]{
\includegraphics[scale=0.22]{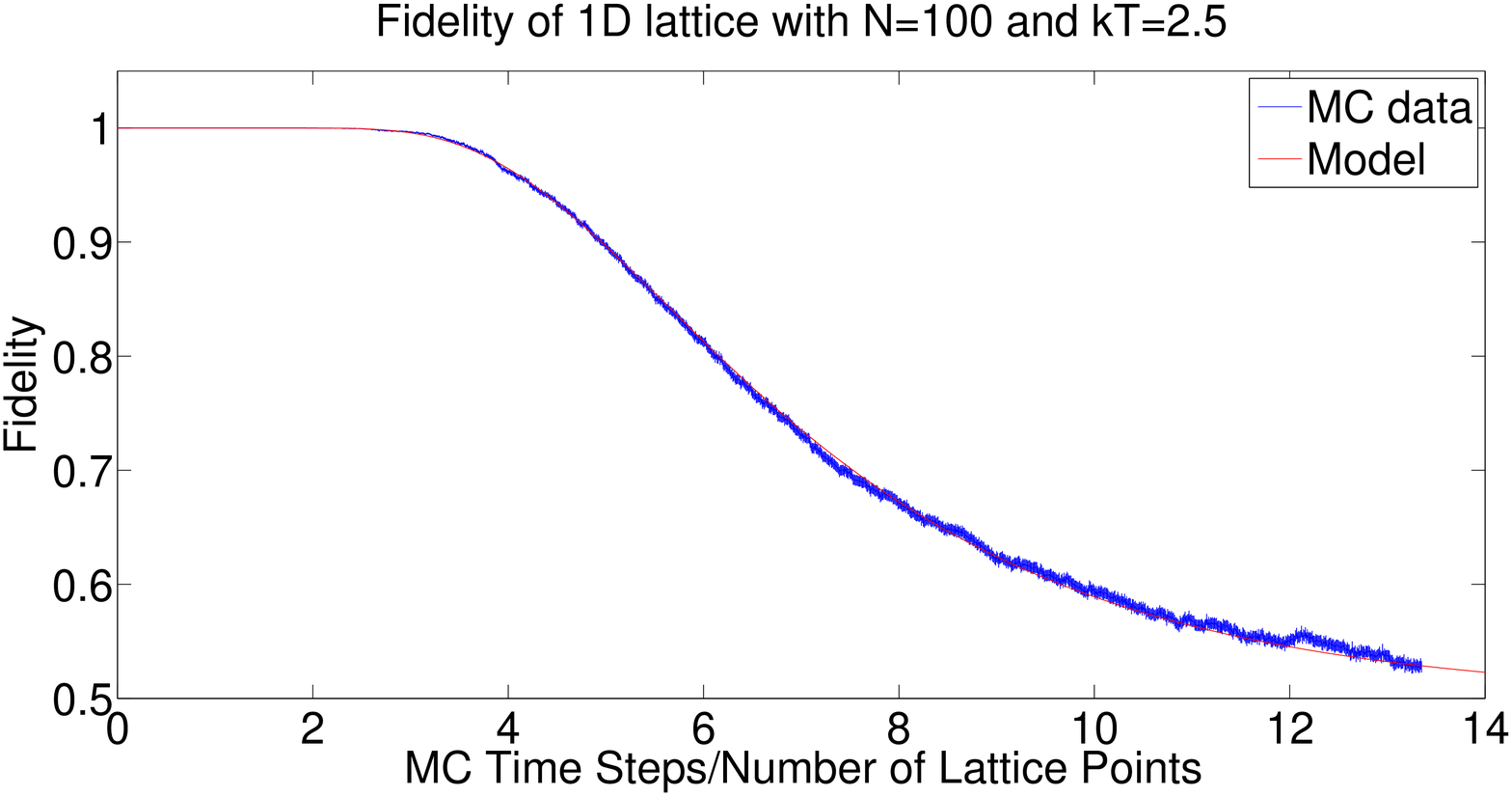}
}
\subfigure[]{
\includegraphics[scale=0.22]{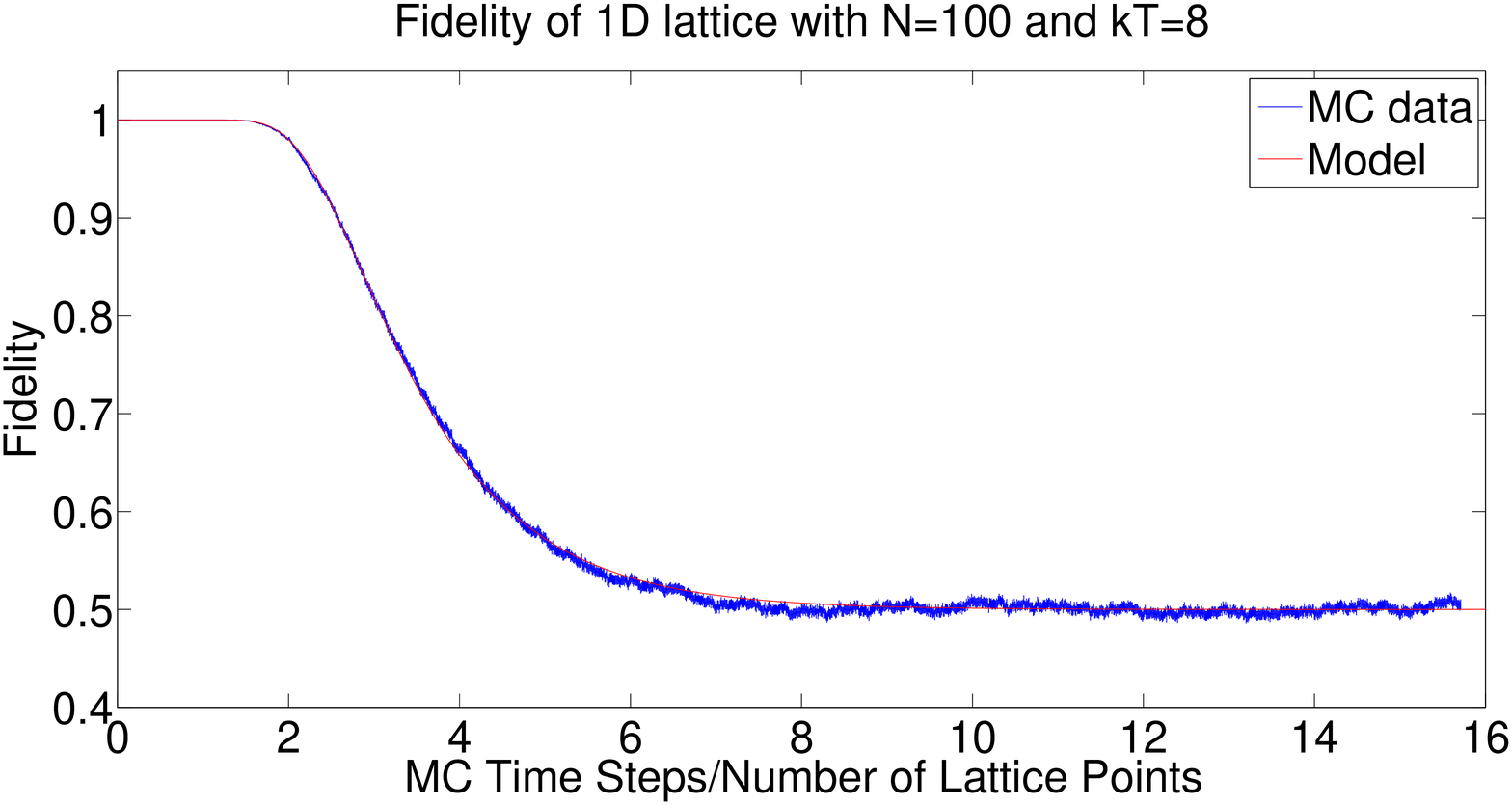}
}
\caption{MC Data (blue with vertical error bars) and model (red) for 1D lattices with (a) $N=100$, $kT=2.5$, $\lambda = 0.1707 \pm 0.0003$, $N_{\rm{eff}} = 46.7 \pm 0.3$, $\chi^{2} = 1093$, $\rm{Degrees\: of\: Freedom} = 1334$ and (b) $N=100$, $kT=8$, $\lambda = 0.4005 \pm 0.0011$, $N_{\rm{eff}} = 99.9 \pm 1.2$, $\chi^{2} = 1421$, $\rm{Degrees\: of\: Freedom} = 1570$}
\end{figure}


\begin{figure} [H]
\subfigure[] {
\includegraphics[scale=0.22]{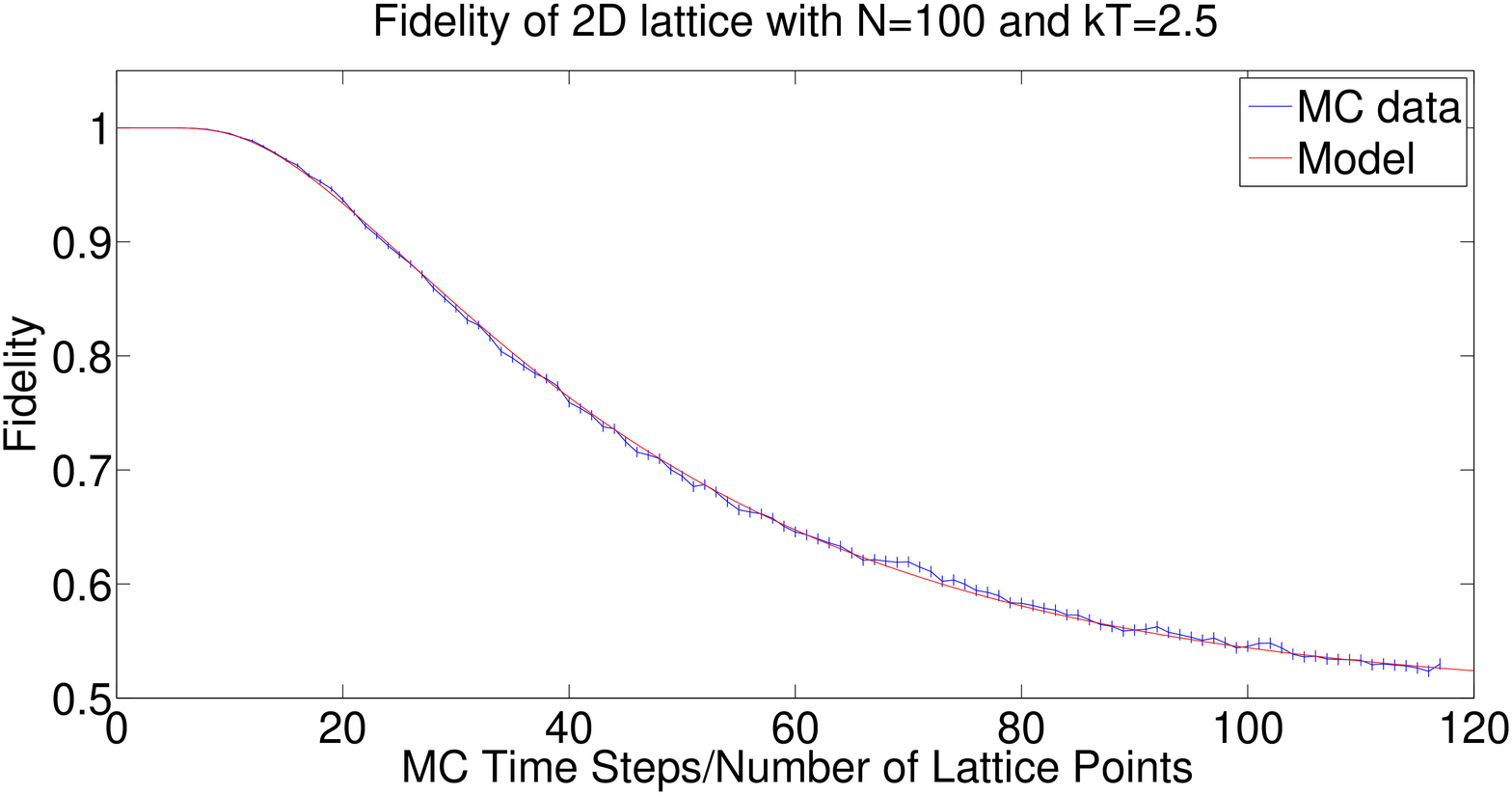}
}
\subfigure[] {
\includegraphics[scale=0.22]{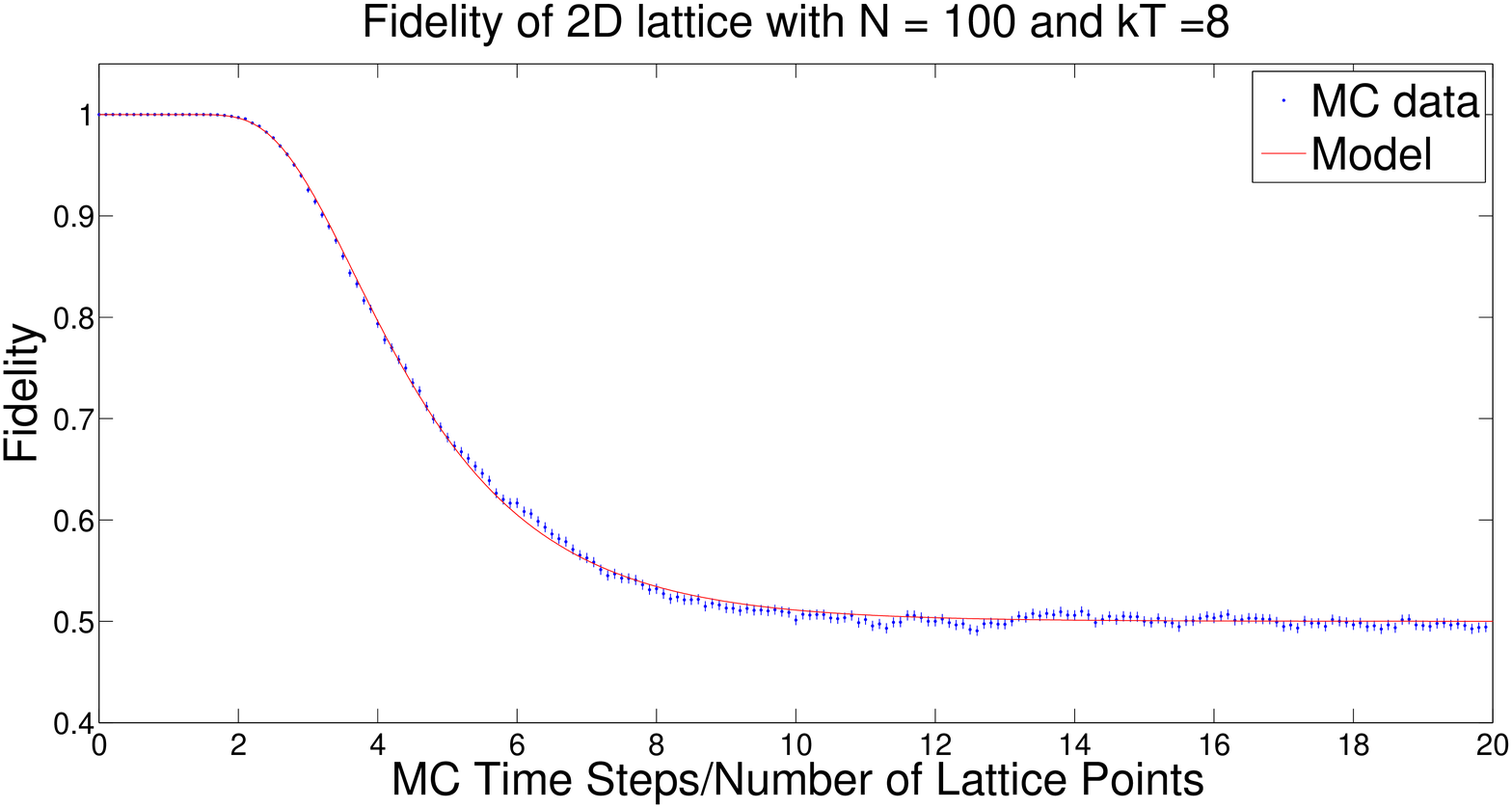}
}
\subfigure[] {
\includegraphics[scale = 0.22] {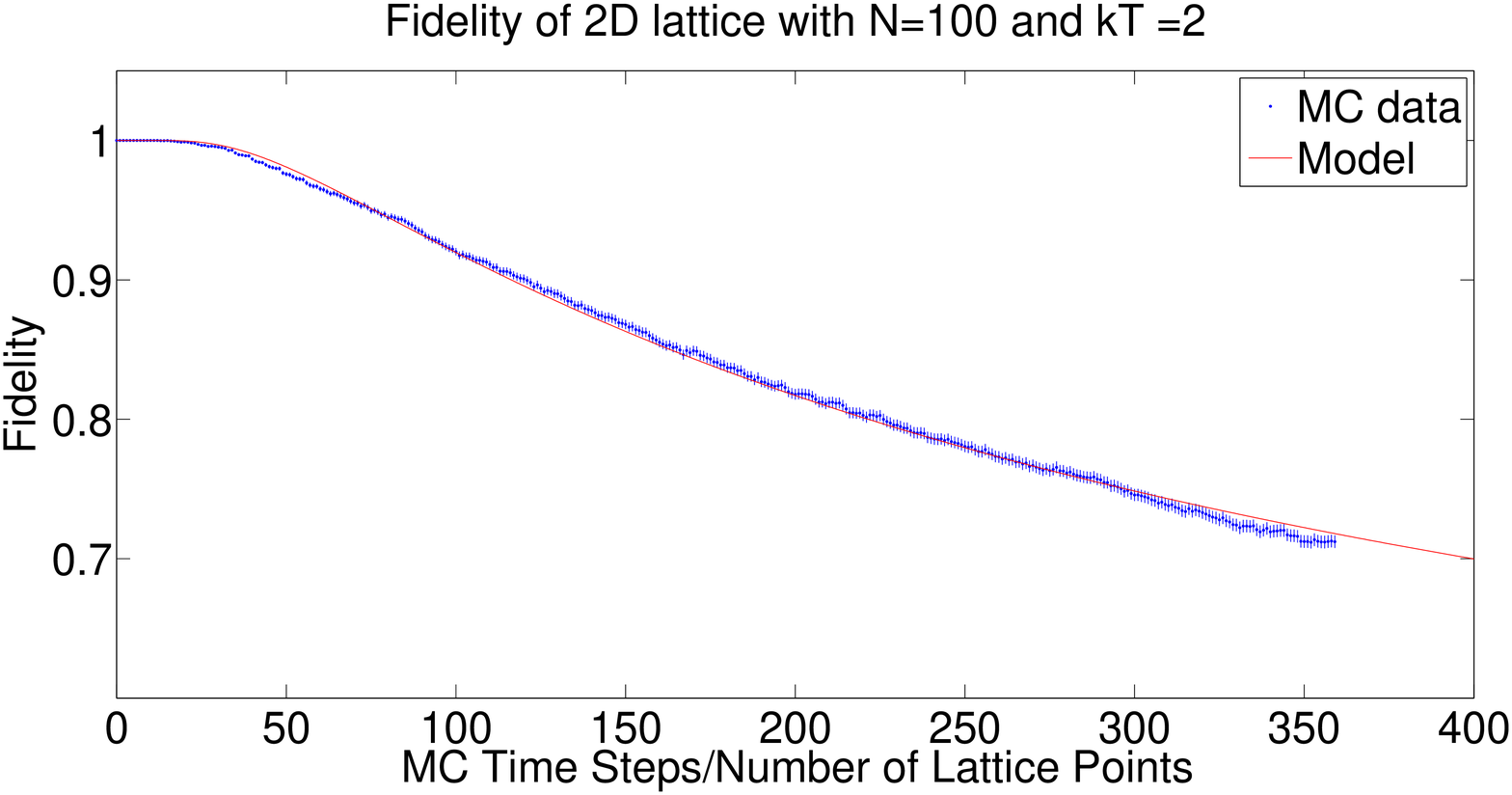}
}
\subfigure[] {
\includegraphics[scale = 0.22] {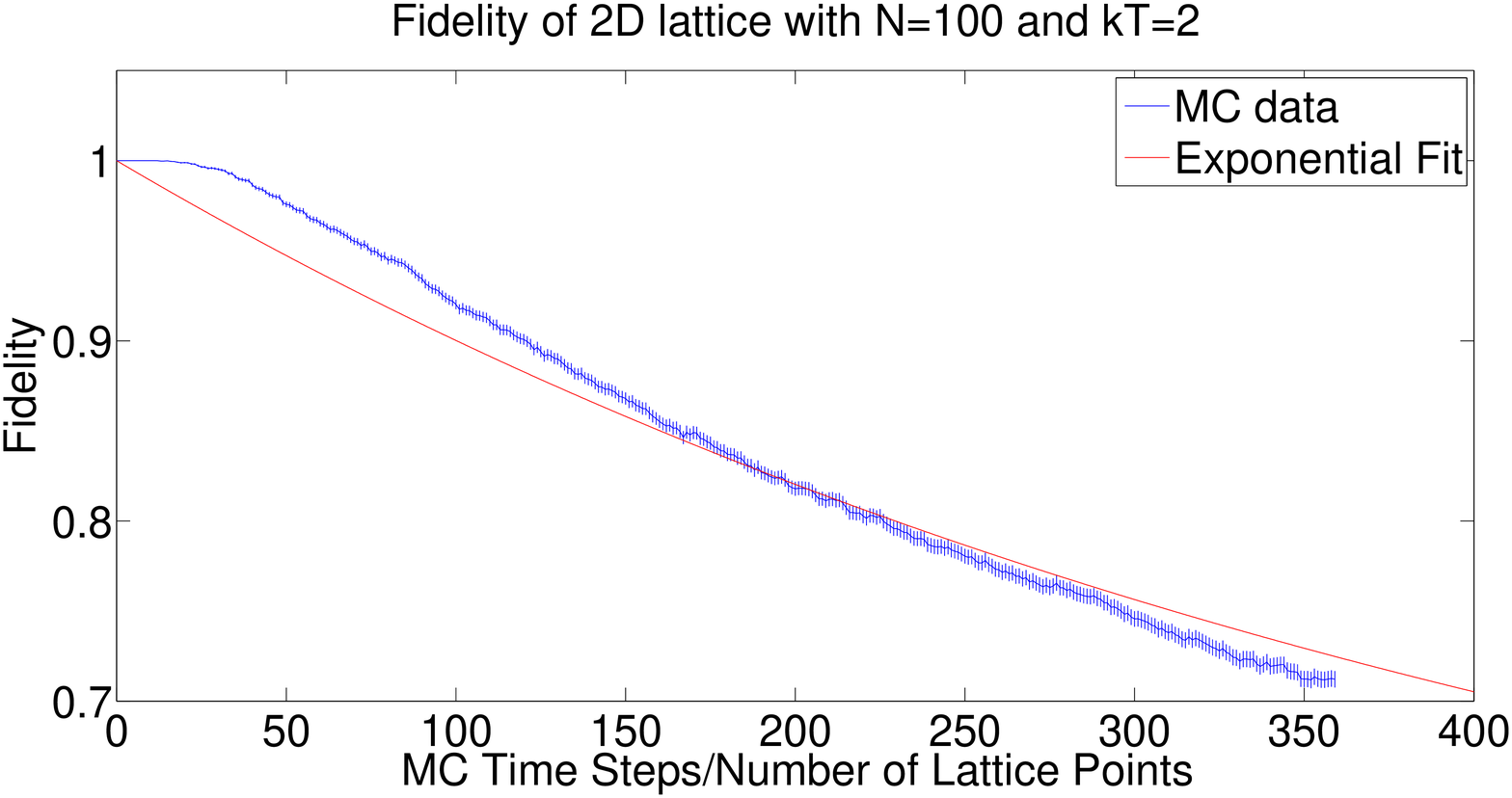}
}
\caption{MC Data (blue with vertical error bars) and model (or exponential fit for (d) only) (red) for 2D lattices with (a) $N=100$, $kT=2.5$, $\lambda = (1.521 \pm 0.010)\times 10^{-2}$, $N_{\rm{eff}} = 5.37 \pm 0.09$, $\chi^{2} = 71$, $\rm{Degrees\: of\: Freedom} = 116$; (b )$N=100$, $kT=8$, $\lambda = 0.282 \pm 0.002$, $N_{\rm{eff}} = 62.2 \pm 1.9$, $\chi^{2} = 198$, $\rm{Degrees\: of\: Freedom} = 198$; (c) $N=100$, $kT = 2$, $\lambda = (8.52 \pm 0.19) \times 10^{-4}$, $N_{\rm{eff}} = 0.80 \pm 0.02$, $\chi^{2} = 684$, $\rm{Degrees\: of\: Freedom} = 358$ and (d) $N=100$, $kT=2$, $\lambda = (1.113 \pm 0.012) \times 10^{-3}$, $\chi^{2} = 434850$, $\rm{Degrees\: of\: Freedom} = 359$.}
\end{figure}

Figures 6a and 7a illustrate one of the key results of this work. For a 1D lattice (figure 6a) $\lambda$ is \emph{independent} of $N$. For a 2D lattice (figure 7a), by contrast, $\lambda$ decreases as $N$ is increased. This is a manifestation of the fundamental difference in the dynamics of the 1D and 2D lattices. In the 1D lattice, no energy input is needed to increase the size of a region of incorrect spin, and hence increasing $N$ does not affect the rate of decay. In the 2D lattice, the opposite is true. Figure 7a also shows that the rate of decrease of $\lambda$ in the 2D lattice falls as we increase $T$. Our data are consistent with an exponential decay of $\lambda$ with increasing $N$ for $T<T_{c}$. Interestingly, and in contrast with the 1D case, for $T>T_{c}$ there does seem to be some reduction in $\lambda$ as $N$ is increased. However, figure 7a indicates that this reduction is subexponential in $N$ for $T>T_{c}$.\\

\begin{figure} [H]
\subfigure[] {
\includegraphics[scale=0.3]{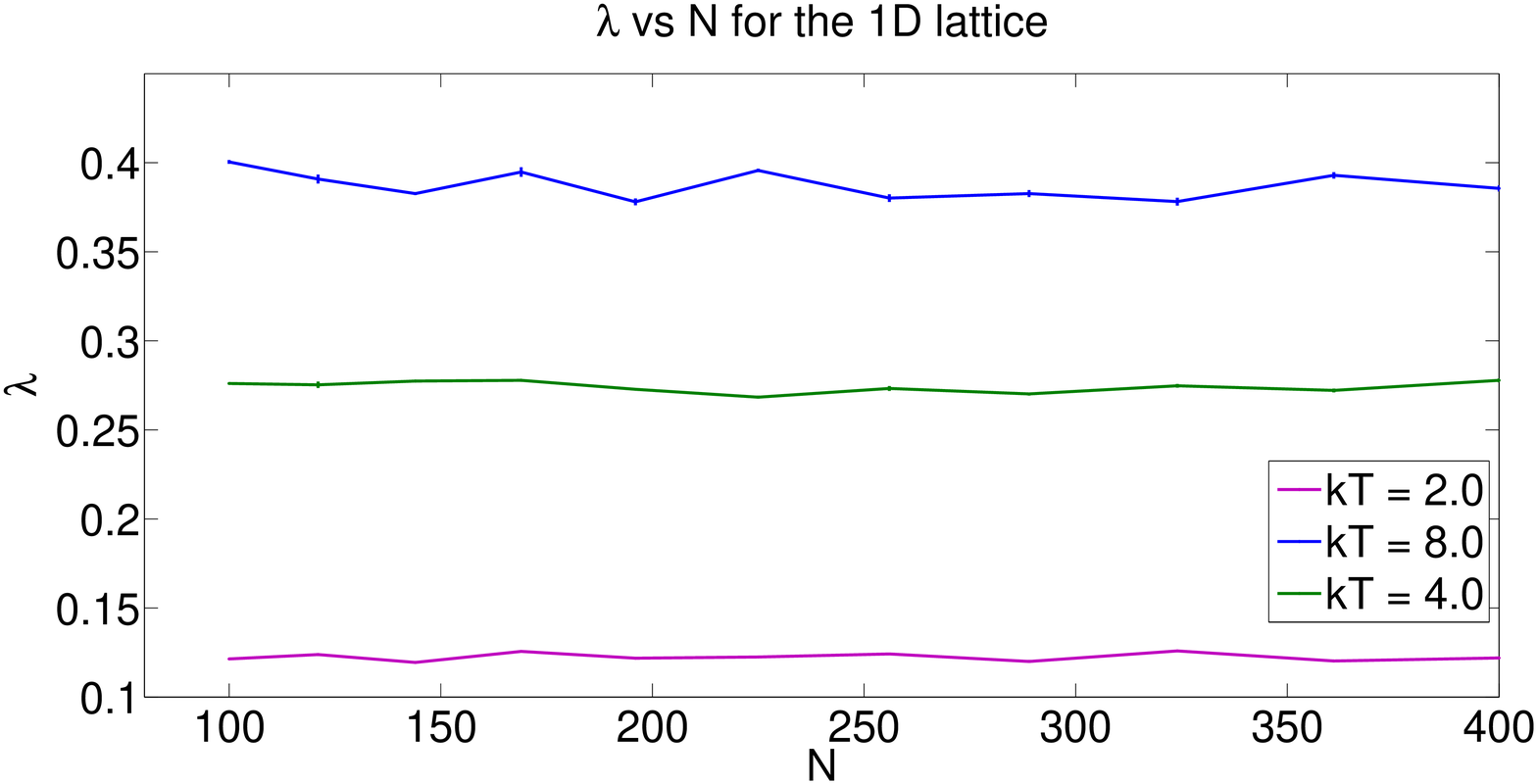}
}
\subfigure [] {
\includegraphics[scale = 0.18]{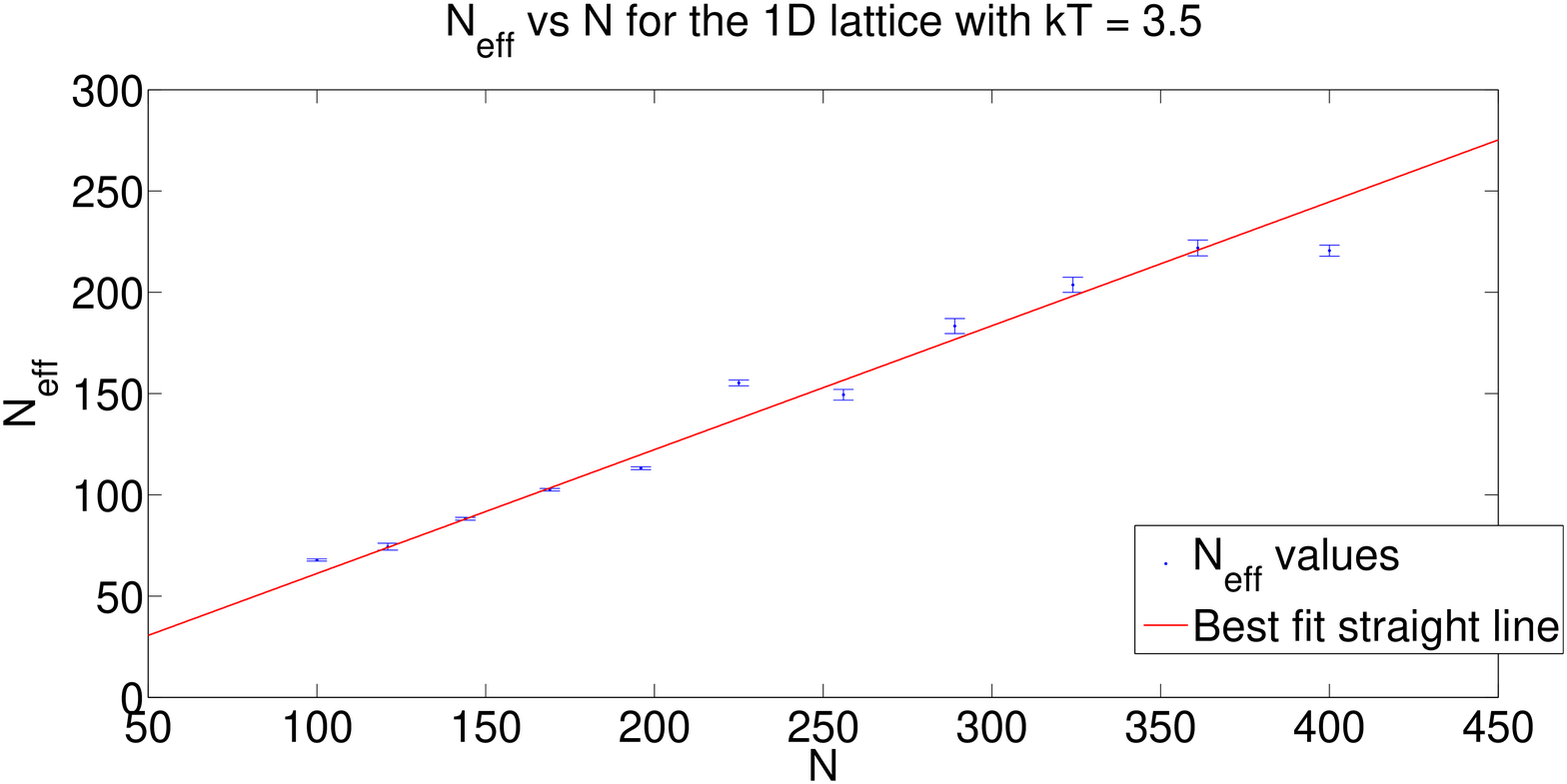}
}
\subfigure [] {
\includegraphics[scale=0.18]{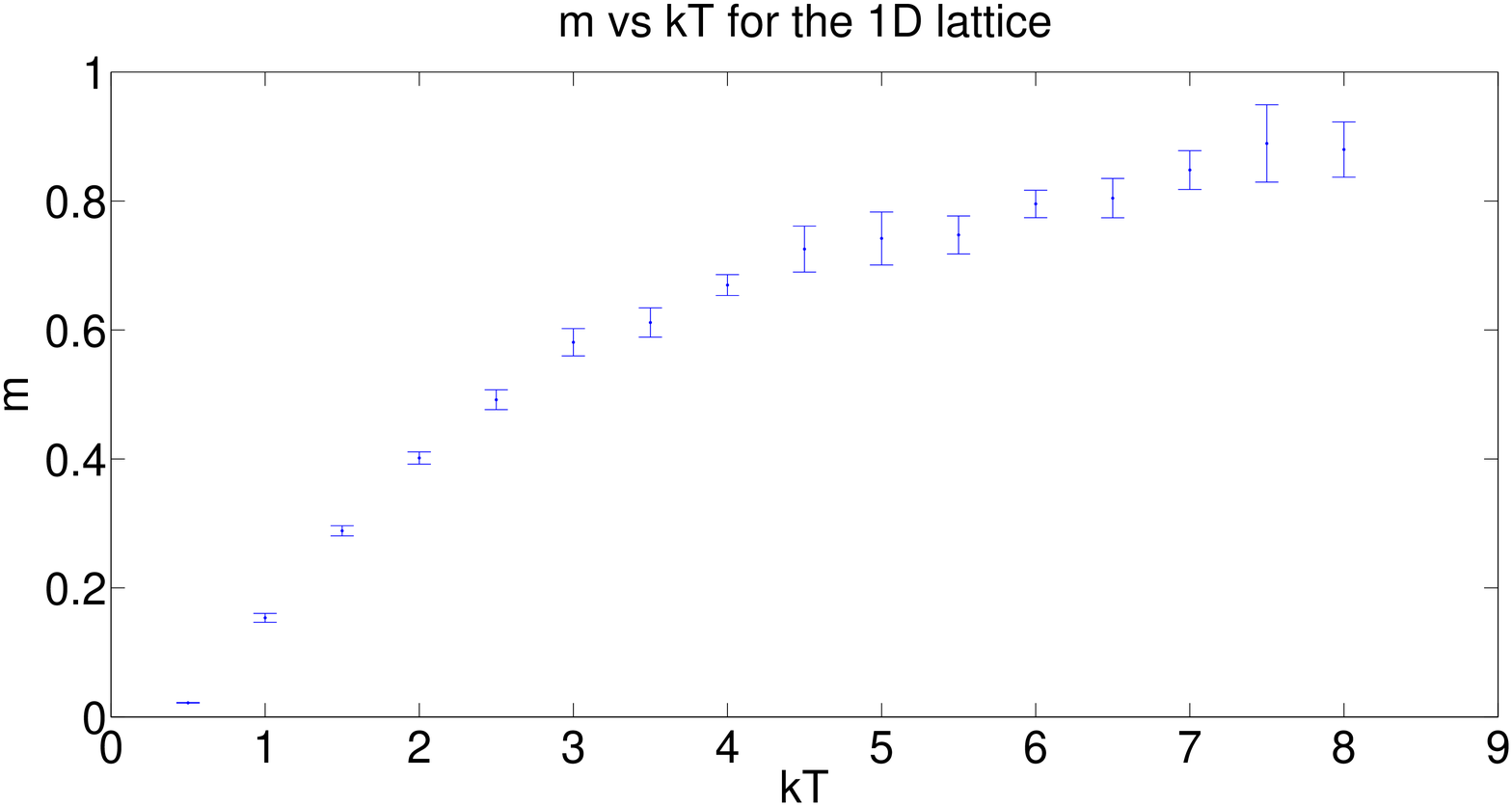}
}
\caption{Variation in (a) $\lambda$ (at $T = {2.0,4.0,8.0}$) and (b) $N_{\rm{eff}}$ (at $T = 3.5$) with $N$ for the 1D lattice. Note some error bars are too small to be visible. We see that $\lambda$ is \emph{independent} of $N$ in this case. (b) shows the fitting $N_{\rm{eff}} = mN$ (red) with $m = 0.61 \pm 0.02$. (c) shows the variation of $m$ with $T$.}
\end{figure}

\begin{figure} [H]
\subfigure[] {
\includegraphics[scale=0.3]{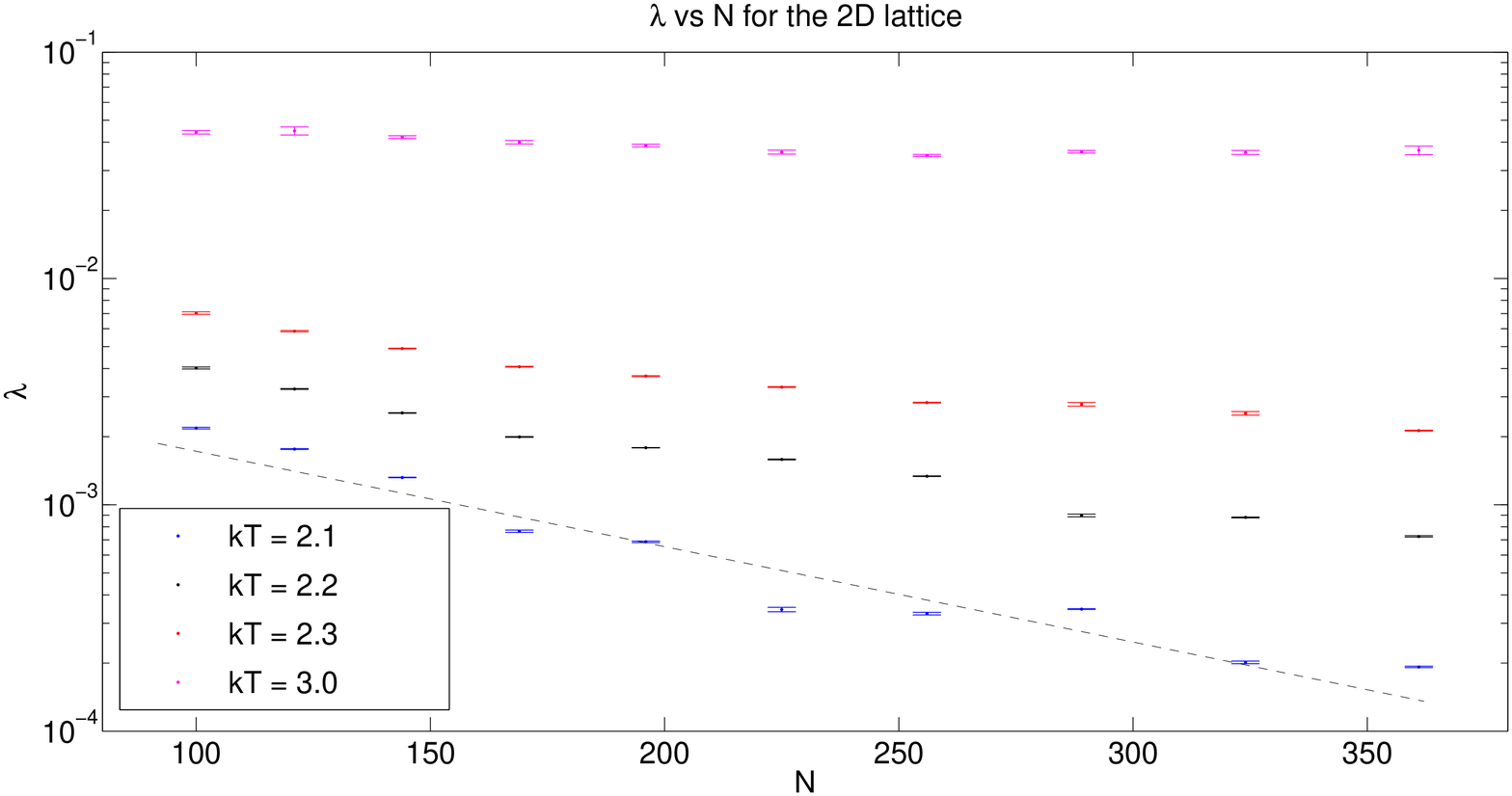}
}
\subfigure[] {
\includegraphics[scale=0.19]{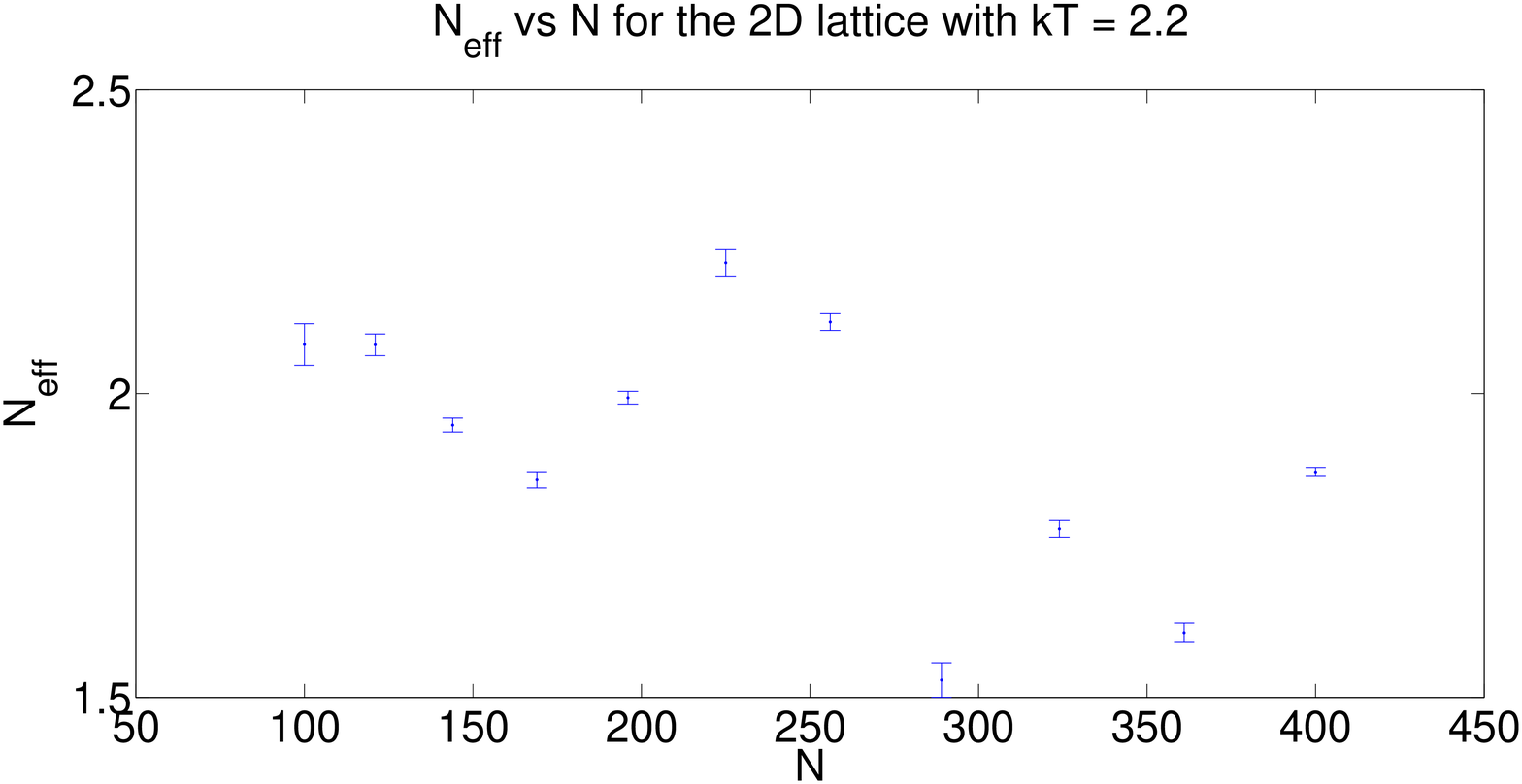}
}
\subfigure[] {
\includegraphics[scale=0.19]{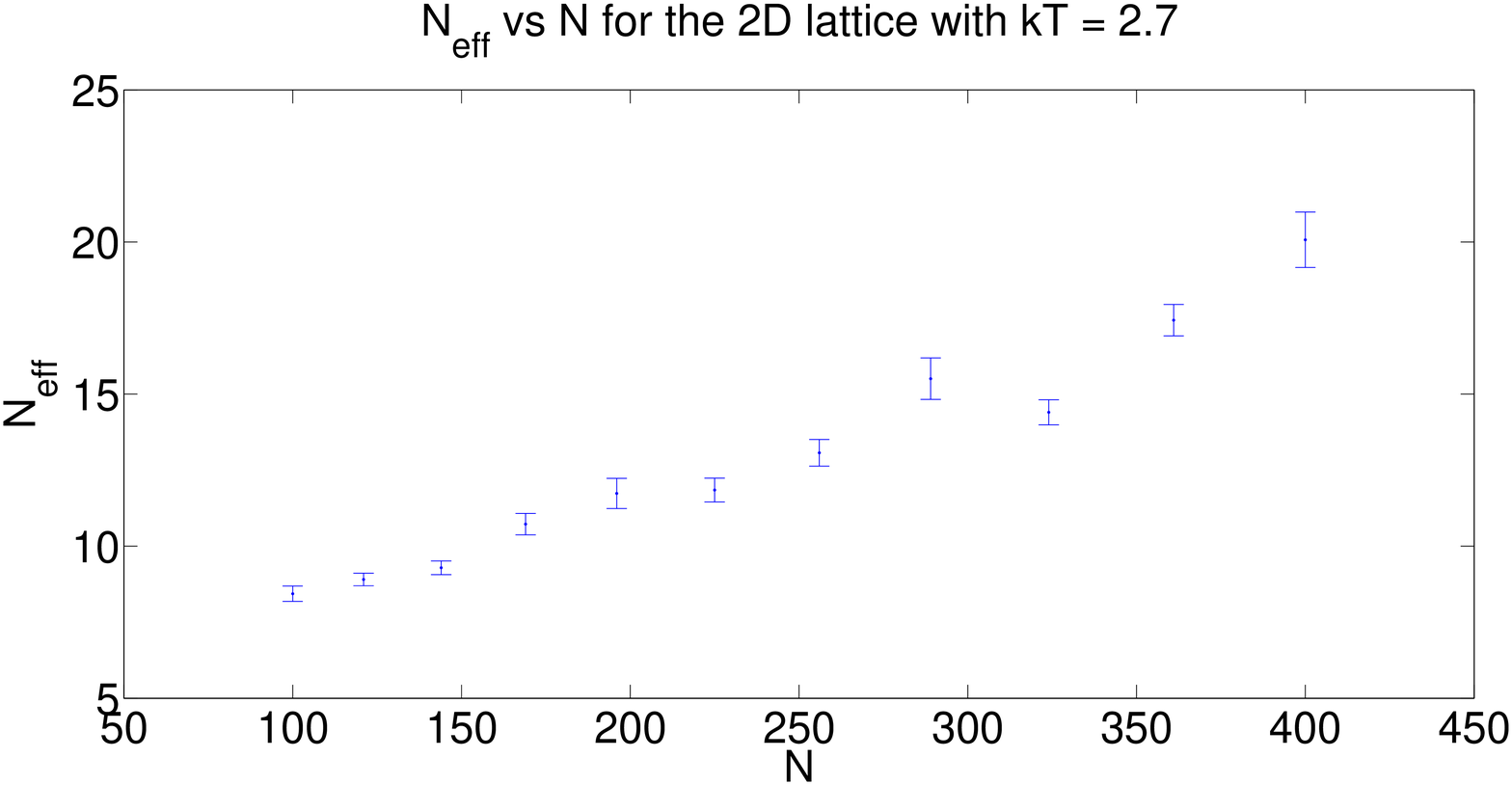}
}
\caption{Variation of (a) $\lambda$ (at $T = {2.1,2.2,2.3,3.0}$, see dashed line to guide the eye) and (b),(c) $N_{\rm{eff}}$ ($T=2.2$ and $T=2.7$) with $N$ for the 2D lattice. These data are consistent with an exponential decrease in $\lambda$ as $N$ increases for $T<T_{c}$, and a sub-exponential decrease for $T>T_{c}$.}
\end{figure}

For a 1D lattice, $N_{\rm{eff}}$ is approximately linearly proportional to $N$: $N_{\rm{eff}} \simeq mN$ (see figure 6b). As $T \rightarrow \infty$, $m \rightarrow 1$ as expected (see figure 6c). Therefore the scaling of the 1D lattice with $N$ is not qualitatively different to that of $N$ non-interacting spins: the only difference is in the value of $m$. The 1D lattice maps directly to the non-interacting system as $T\rightarrow \infty$.\\

There are two competing factors influencing the variation of $N_{\rm{eff}}$ as $N$ is increased in the 2D lattice. Firstly, a larger lattice can accommodate more spin regions of a given size, which would lead to a positive correlation between $N_{\rm{eff}}$ and $N$. Secondly, a larger lattice has more inherent stability due to interactions between spins, leading to to a negative correlation between $N_{\rm{eff}}$ and $N$. Our simulations show that, just below the critical temperature, $N_{\rm{eff}}$ and $N$ are approximately negatively correlated (figure 7b), whereas, for $T>T_{c}$, $N_{\rm{eff}}$ and $N$ are approximately positively correlated (figure 7c). Thus the phase transition appears to emerge naturally from the proposed model in the form of the variation of both $\lambda$ and $N_{\rm{eff}}$ with $N$. We also notice that these correlations are not very strong for low $T$, indicating that the model is breaking down here. This is likely due to both the Gaussian approximation becoming poor, and the model becoming physically less applicable.

\section{Conclusions and Further Work}

Our model offers an intuitive view of the evolution of the Ising lattice at high and intermediate temperatures and provides continuity between the interacting and non-interacting cases: as we take the temperature to infinity, the model parameters approach their non-interacting values. The work described here entails a proof of principle of the proposed model, and an exploration of what we can infer about the 1D and 2D Ising lattices from its parameters. Our model gives a quantitative backing to the commonly held belief that at low temperatures the 2D Ising lattice is a good physical memory, whereas the 1D Ising lattice is not. More specifically, in the 2D case $\lambda $ decreases as we increase the system size, giving us significant returns on our investment in extra spins. In the 1D case, by contrast, $\lambda$ is independent of system size so we get very little return on an investment in extra spins. Furthermore, we have seen the phase transition of the 2D lattice emerge from our model, as the decrease in $\lambda$ as $N$ increases appears to change from exponential to sub-exponential as we move from $T<T_{c}$ to $T>T_{c}$.

Both classical and quantum data storage are likely to be of theoretical and practical interest in future years, prompting the study of the fidelity of a range of physical memories. This article begins this process with the study of the Ising lattice, the archetypal classical physical memory.

A natural extension of this work would consider the applicability of the new model for $h\neq0$, for $D > 2$, and for different boundary conditions. The low temperature case, for which neither the model proposed here nor an exponential model is entirely satisfactory, also merits further study.

\ack
This work was carried out as part of the Undergraduate Research Opportunities Programme at Imperial College. Both  FVD and SDB were supported via SDB's Royal Society University Research Fellowship.

\appendix
\section{Variation of Model Parameters with $T$}

The variation in the model parameters with $T$ are qualitatively similar for the 1D and 2D lattices. As $T$ is increased, $\lambda \rightarrow 0.5$ and $N_{\rm{eff}} \rightarrow N $ as expected (see figure A1). The primary advantage of the 2D lattice is therefore in the scaling of the fidelity with $N$.\\

\begin{figure} [H]
\subfigure[] {
\includegraphics[scale=0.22]{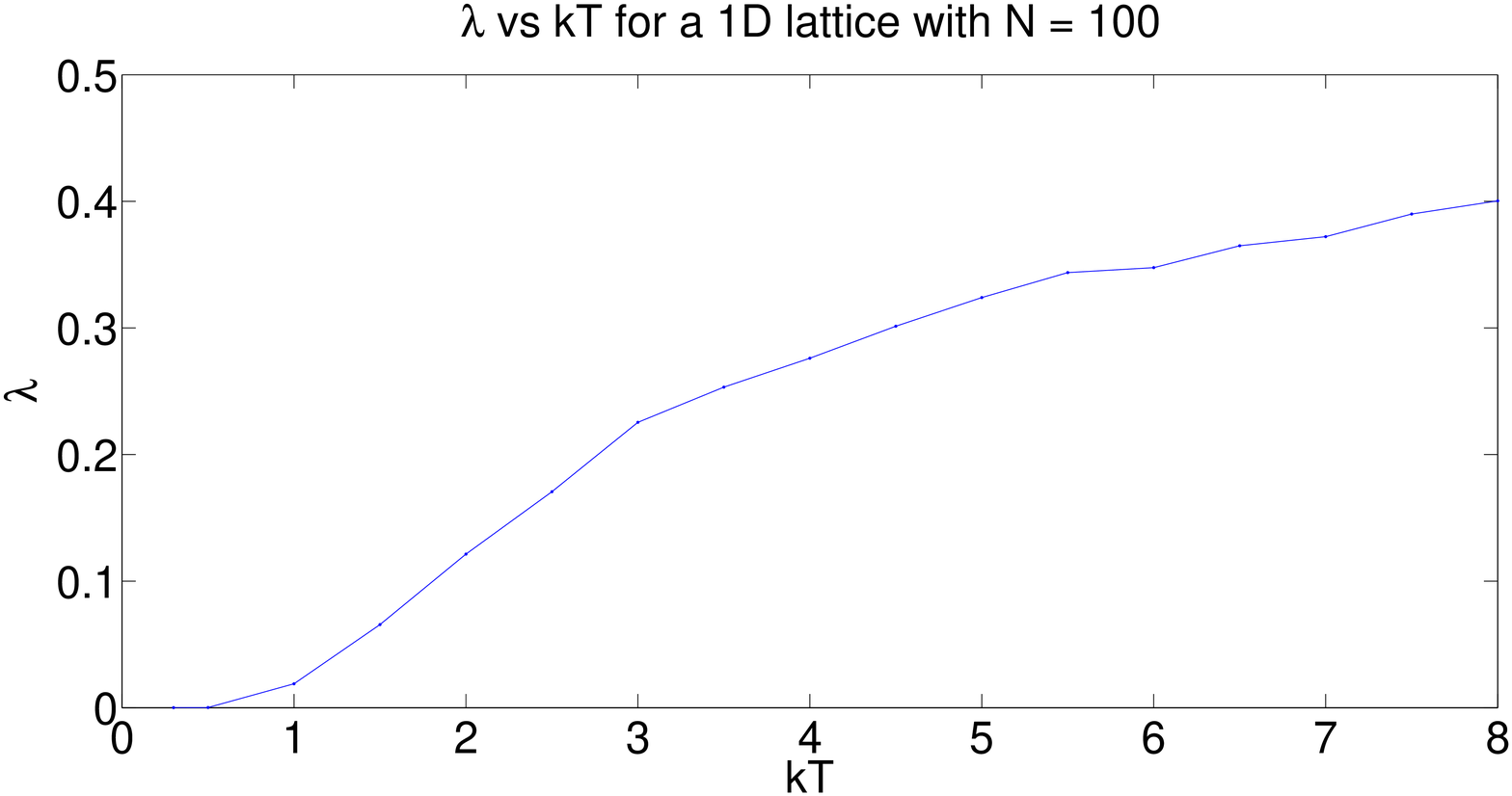}
}
\subfigure[] {
\includegraphics[scale=0.22]{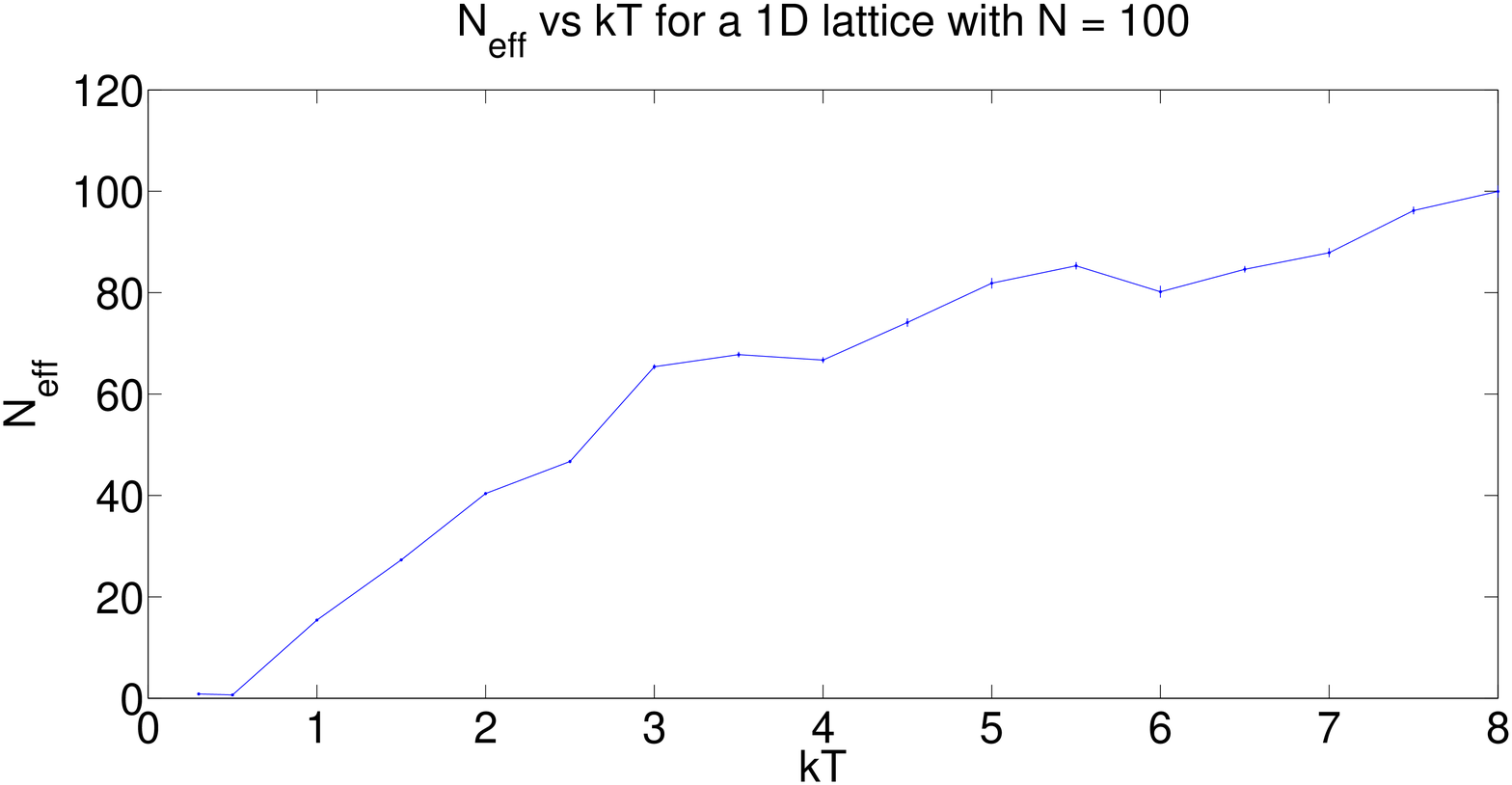}
}
\subfigure[] {
\includegraphics[scale=0.22]{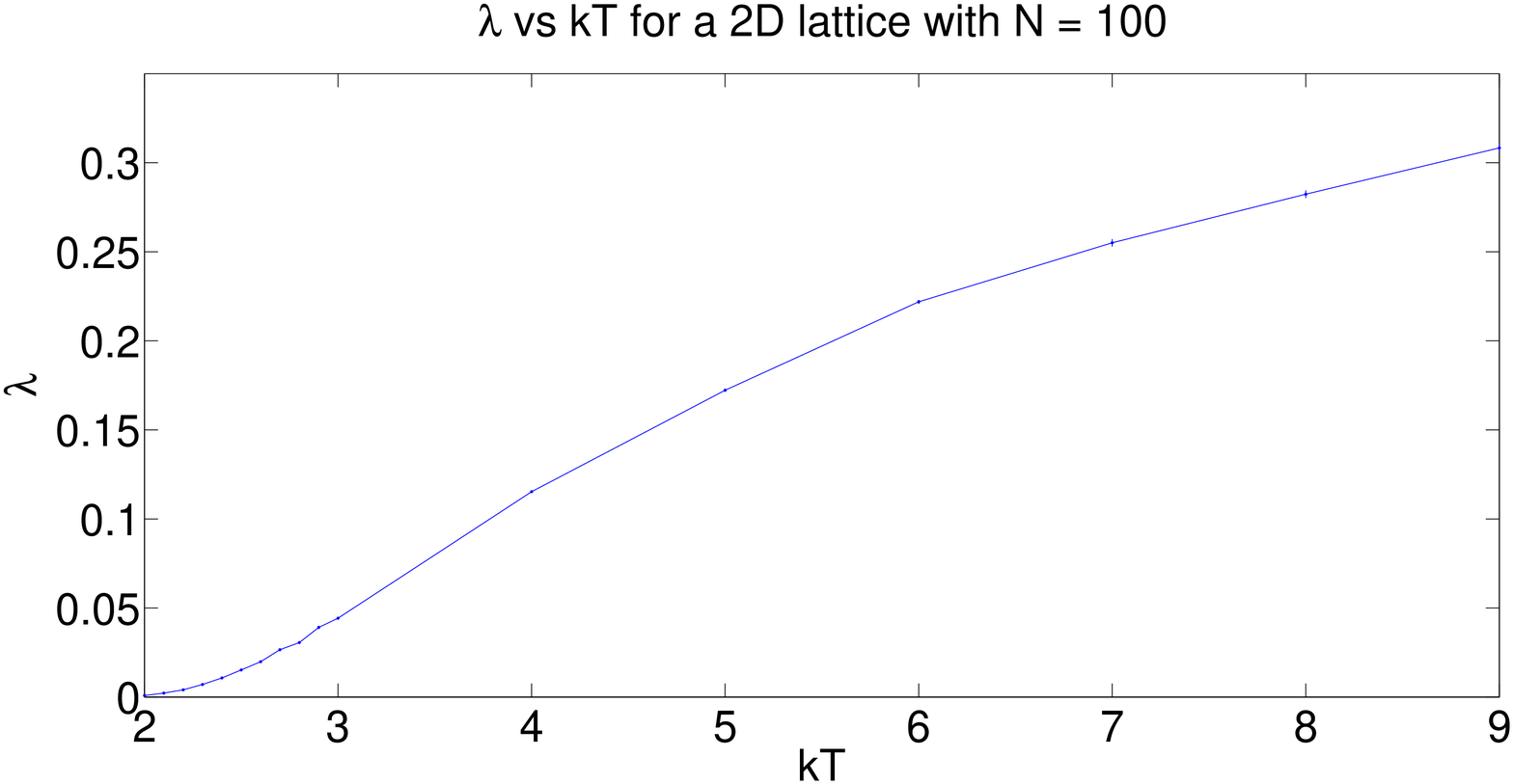}
}
\subfigure[] {
\includegraphics[scale=0.22]{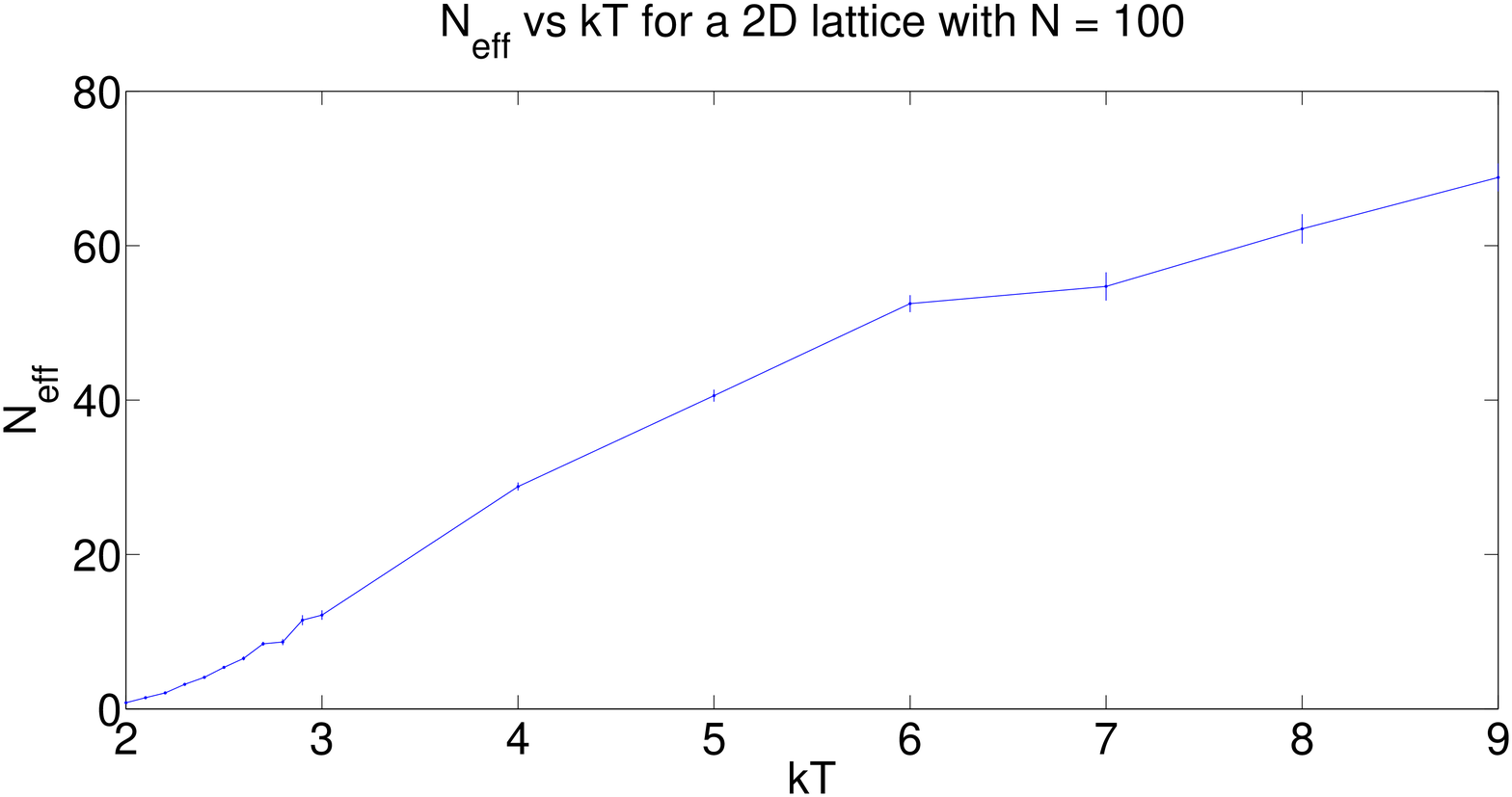}
}
\caption{Variation in (a),(c) $\lambda$ and (b),(d) $N_{\rm{eff}}$ with $kT$ for the 1D and 2D lattices with $N = 100$. Note some error bars are too small to be visible.}
\end{figure}

\section*{References}

\bibliographystyle{unsrt}
\bibliography{Ising}

\end{document}